\documentclass[10pt,conference]{IEEEtran}
\IEEEoverridecommandlockouts
\usepackage{cite}
\usepackage{amsmath,amssymb,amsfonts}
\usepackage{graphicx}
\usepackage{textcomp}
\usepackage{xcolor}
\usepackage{xfrac}
\usepackage{hyperref}
\usepackage{listings}
\usepackage{tcolorbox}
\usepackage{paralist} 
\usepackage[hang,flushmargin]{footmisc}

\newcommand{\sysname}{Rango\xspace}
\newcommand{\datasetname}{CoqStoq\xspace}

\usepackage{xcolor}

\definecolor{ktcolor}{RGB}{0, 100, 0}
\definecolor{efcolor}{RGB}{100, 50, 255}
\definecolor{slcolor}{RGB}{255, 50, 50}
\definecolor{nscolor}{RGB}{255, 0, 255}
\definecolor{pccolor}{RGB}{255, 140, 0}
\definecolor{jfcolor}{RGB}{0, 0, 170}
\definecolor{ybcolor}{RGB}{100, 175, 170}
\definecolor{ascolor}{RGB}{175, 170, 0}
\definecolor{kfcolor}{RGB}{0, 170, 175}

\newcommand\blfootnote[1]{%
  \begingroup
  \renewcommand\thefootnote{}\footnote{#1}%
  \addtocounter{footnote}{-1}%
  \endgroup
}

\definecolor{codegreen}{rgb}{0,0.6,0}
\lstdefinestyle{CoqStyle}{
    backgroundcolor=\color{white},   
    commentstyle=\color{codegreen},
    keywordstyle=\color{blue},
    numberstyle=\tiny\color{black},
    stringstyle=\color{red},
    basicstyle=\ttfamily\scriptsize,
    morekeywords={Require,Import,Proof,Qed,Lemma,Theorem,Admitted},
    breakatwhitespace=false,         
    breaklines=true,                 
    captionpos=b,                    
    keepspaces=true,                 
    numbers=left,                    
    xleftmargin=5.0ex,
    numbersep=5pt,                  
    showspaces=false,                
    showstringspaces=false,
    showtabs=false,                  
    tabsize=2
}

\def\BibTeX{{\rm B\kern-.05em{\sc i\kern-.025em b}\kern-.08em
    T\kern-.1667em\lower.7ex\hbox{E}\kern-.125emX}}
\usepackage{algorithm}
\usepackage[noend]{algpseudocode}
\usepackage{amsmath}
\usepackage{amsfonts}
\usepackage{amsthm}
\usepackage{balance}
\usepackage{graphicx}
\usepackage{xspace}
\usepackage{pifont}
\usepackage{multirow}
\usepackage{array}
\usepackage{booktabs}
\usepackage{microtype}
\usepackage{enumitem}
\usepackage{color,colortbl}
\usepackage{dblfloatfix}
\usepackage{framed}
\usepackage{subfigure}
\usepackage{xtab,afterpage}
\usepackage{tcolorbox}

\usepackage{listings}

\definecolor{darkgreen}{rgb}{0,0.42,0.24}
\lstdefinelanguage{diff}{
	morecomment=[f][\color{blue}]{@@},     
	morecomment=[f][\color{red}]-,         
	morecomment=[f][\color{darkgreen}]+,       
	morecomment=[f][\color{magenta}]{---}, 
	morecomment=[f][\color{magenta}]{+++},
}

\newcommand\lt[1]{{\lstinline+#1+}} 
\renewcommand\t[1]{{\lstinline+#1+}} 
\definecolor{dkgreen}{rgb}{0,0.5,0}
\definecolor{dkred}{rgb}{0.5,0,0}
\definecolor{gray}{rgb}{0.5,0.5,0.5}
\let\origthelstnumber\thelstnumber
\makeatletter
\newcommand*\Suppressnumber{%
  \lst@AddToHook{OnNewLine}{%
    \let\thelstnumber\relax%
     \advance\c@lstnumber-\@ne\relax%
    }%
}

\newcommand*\Reactivatenumber{%
  \lst@AddToHook{OnNewLine}{%
   \let\thelstnumber\origthelstnumber%
   \advance\c@lstnumber\@ne\relax}%
}

\definecolor{LightGray}{gray}{0.9}
\definecolor{Gray}{gray}{0.8}

\usepackage{tikz}

\usepackage{hyperref}
\hypersetup{%
 pdftitle = {Rango: Adaptive Retrieval-Augmented Proving for Automated Software Verification},
 pdfkeywords = {},
 pdfauthor = {Kyle Thompson, Nuno Saavedra, Pedro Carrott, Kevin Fisher, Alex Sanchez-Stern, Yuriy Brun, Jo\~{a}o F. Ferreira, Sorin Lerner, and Emily First},
 bookmarksnumbered,
 bookmarksopen=false,
 urlcolor=black,
 linkcolor=black,
 citecolor=black,
 colorlinks=true,
 pdfstartview={FitH},
}

\begin{document}

\title{Rango: Adaptive Retrieval-Augmented Proving for Automated Software Verification}

\author{
\IEEEauthorblockN{\hspace{3em}Kyle Thompson}
\IEEEauthorblockA{
\textit{\hspace{4em}University of California\hspace{1em}}\\
\hspace{3em}San Diego, CA, USA \\
\hspace{3em}r7thompson@ucsd.edu}
\and
\IEEEauthorblockN{Nuno Saavedra}
\IEEEauthorblockA{
\textit{\hspace{2em}INESC-ID \& IST, University of Lisbon\hspace{2em}}\\
Lisbon, Portugal \\
nuno.saavedra@tecnico.ulisboa.pt}
\and
\IEEEauthorblockN{Pedro Carrott}
\IEEEauthorblockA{
\textit{\hspace{2em}Imperial College London\hspace{2em}}\\
London, UK \\
pedro.carrott@imperial.ac.uk}
\and
\IEEEauthorblockN{\hspace{3em}Kevin Fisher}
\IEEEauthorblockA{
\textit{\hspace{5em}University of California\hspace{2em}}\\
\hspace{3em}San Diego, CA, USA \\
\hspace{3em}kfisher@ucsd.edu}
\and
\IEEEauthorblockN{Alex Sanchez-Stern}
\IEEEauthorblockA{
\textit{\hspace{2em}University of Massachusetts\hspace{2em}}\\
Amherst, MA, USA \\
sanchezstern@cs.umass.edu} 
\and
\IEEEauthorblockN{Yuriy Brun}
\IEEEauthorblockA{
\textit{\hspace{2em}University of Massachusetts\hspace{2em}}\\
Amherst, MA, USA \\
brun@cs.umass.edu}
\and
\IEEEauthorblockN{Jo\~{a}o F. Ferreira}
\IEEEauthorblockA{
\textit{\hspace{2em}INESC-ID \& IST, University of Lisbon\hspace{2em}}\\
Lisbon, Portugal \\
joao@joaoff.com}
\and
\IEEEauthorblockN{Sorin Lerner\hspace{4em}}
\IEEEauthorblockA{
\textit{University of California\hspace{4em}}\\
San Diego, CA, USA\hspace{4em} \\
lerner@cs.ucsd.edu\hspace{4em}}
\and
\IEEEauthorblockN{Emily First}
\IEEEauthorblockA{
\textit{\hspace{2em}University of California\hspace{2em}}\\
San Diego, CA, USA \\
emfirst@ucsd.edu}
}

\maketitle

\thispagestyle{plain}
\pagestyle{plain}

\begin{abstract}
Formal verification using proof assistants, such as Coq, enables the creation of high-quality software.
However, the verification process requires significant expertise and manual effort to write proofs.
Recent work has explored automating proof synthesis using machine learning and large language models (LLMs).
This work has shown that identifying relevant premises, such as lemmas and definitions, can aid synthesis.
We present \sysname, a fully automated proof synthesis tool for Coq that automatically identifies relevant premises and also similar proofs from the current project and uses them during synthesis.  
\sysname uses retrieval augmentation at every step of the proof to automatically determine which proofs and premises to include in the context of its fine-tuned LLM. 
In this way, \sysname adapts to the project and to the evolving state of the proof. 
We create a new dataset, \datasetname, of 2,226 open-source Coq projects and 196,929 theorems from GitHub, which includes both training data and a curated evaluation benchmark of well-maintained projects.
On this benchmark, \sysname synthesizes proofs for 32.0\% of the theorems, which is 29\% more theorems than the prior state-of-the-art tool Tactician.
Our evaluation also shows that \sysname adding relevant proofs to its context leads to a 47\% increase in the number of theorems proven. 
\blfootnote{\noindent\fbox{\parbox{.98\columnwidth}{Kyle Thompson, Nuno Saavedra, Pedro Carrott, Kevin Fisher, Alex
  Sanchez-Stern, Yuriy Brun, Jo{\~{a}}o F. Ferreira, Sorin Lerner, and Emily First.
  Rango: Adaptive Retrieval-Augmented Proving for Automated Software Verification.
  In \emph{Proceedings of the 47th International Conference on Software
  Engineering (ICSE)}, Ottowa, ON, Canada, April 2025.}}}
\end{abstract}


\section{Introduction}
\label{sec:intro}

The cost of poor software quality is alarmingly high, with estimates suggesting that it incurs trillions of dollars in losses annually in the United States alone~\cite{Krasner2021}.  
Formal verification, which enables developers to mathematically prove that software adheres to its intended behaviors and specifications, has been shown to help improve software quality. 
Notably, a study investigating C compilers~\cite{Yang11a}, among them CompCert~\cite{Leroy06}, LLVM, and GCC, observed that the only compiler for which no bugs were found was CompCert~\cite{Yang11a}, which is formally verified using the Coq proof assistant.

\looseness-1
While formal verification can lead to high-quality software, it is expensive. 
For example, the code required to verify CompCert is 8 times as large as the code implementing its functionality~\cite{Leroy06}. 
A promising line of work towards automating formal verification is to automatically generate the proofs 
using machine learning techniques~\cite{First20oopsla, First22icse, Sanchez-Stern23toplas, Sanchez20, Blaauwbroek20, Blaauwbroek24ICML}.
Within this research space, there has been a recent and exciting line of work on exploring large language models (LLMs) for proof generation~\cite{Polu20, Jiang21, Han21, Jiang22, Yang23}.

Prior LLM-based approaches for proof automation have shown that \emph{premises}, such as lemmas and definitions, are important to add to the context~\cite{Yang23, mikuła2023magnushammer}. This builds off of \emph{retrieval-augmented generation (RAG)}~\cite{Lewis2020} approaches, which use a separate search step to add context to an LLM. For the task of proof synthesis, we call this technique \emph{retrieval-augmented proving (RAP)}, where a separate search step retrieves relevant information for proving a given theorem. One limitation of prior approaches using RAP is that they do not fully exploit the local information that is available when synthesizing proofs.

In this paper, we show that an essential component of RAP is to \emph{also} include similar proofs in the context --
not only at the beginning of the proof, but to continue to give the LLM sources of inspiration and knowledge, adapting to the evolving proof state. 
The intuition is that having similar proofs in the context of the LLM, as determined at each step in the proof, can help guide the LLM in the right direction.
Our work also distinguishes itself within the broader scope of all machine-learning-based proof generation approaches by using \emph{both} premises and proofs in an online setting.

We reify this idea of adding relevant proofs, not just premises, to the context of an LLM in an approach and tool called \sysname.
At each proof step, \sysname first determines which proofs and premises from the current project are most relevant for generating the next step.
By retrieving in-project information, \sysname is able to learn local proof strategies, adapting itself to the current project.
Then, the next step is generated using a language model where the most relevant proofs and premises are given as context in addition to the current proof script (the steps taken so far) and the proof state (a set of logical formulas describing the goals that remain to be proven). Furthermore, the assessment of which proofs are relevant is done \emph{at each step in the proof}, thus being able to adapt to the evolving nature of the proof.

To train \sysname effectively, we collected a new large corpus of Coq data, which we call \datasetname. We mine this corpus from GitHub using the CoqPyt Python client for CoqLSP~\cite{carrott2024coqpyt}. We split \datasetname into training data and a benchmark, on which we evaluate \sysname.
The benchmark contains all of the projects from a previous benchmark, CoqGym~\cite{Yang19}, that compile in Coq 8.18.  
Additionally, it contains CompCert and four projects from the Coq Community repository that are committed to long-term maintenance~\cite{CoqCommunity2024}. 


In summary, this paper's main contributions are:

\begin{itemize}

  \item An approach, \sysname, to synthesizing proofs that adds to the context of the LLM not just premises but also relevant proofs. In this way, \sysname adapts to the project \emph{and} to the proof state at each step.
  
  \item A new dataset, \datasetname, for proof synthesis in Coq, comprising 196,929 theorems, and 2,225,515 proof steps from 2,226 different GitHub repositories. The dataset is split into training data and an evaluation benchmark.

  \item An evaluation on \datasetname comparing \sysname to three state-of-the-art systems, Proverbot9001~\cite{Sanchez20}, Tactician~\cite{Blaauwbroek20}, and Graph2Tac~\cite{Blaauwbroek24ICML}. Our evaluation shows that \sysname does better than these tools, by proving 29\% more theorems than Tactician, 66\% more than Proverbot9001 and 4\% more than Graph2Tac. Our evaluation also shows that adding relevant proofs to the context in \sysname is important, leading to a 47\% increase in the number of theorems proven.

\end{itemize}
We release \sysname, \datasetname, all trained models appearing in this paper, and all of the code required to reproduce the experiments in this paper at this link: \url{https://github.com/rkthomps/coq-modeling}.



\section{Motivating Example in Coq}
\label{sec:background}


To motivate our approach, we explain how formal verification works in Coq, and then demonstrate through a real example how \sysname helps automate the proof-writing process. 

In Coq, a proof engineer can state a theorem about their code and then write a proof that the theorem holds true. Theorems in Coq are types, and so proving them true amounts to constructing a \emph{proof term} of the same type in Coq's Gallina language. However, since writing that term directly is challenging, proof engineers typically write \emph{proof scripts} in Coq's Ltac language, which consist of \emph{tactics}, such as \texttt{induction}. When executed, these tactics guide Coq in the construction of a complete proof term. Coq provides feedback after each tactic application and displays the current \emph{proof state}, which includes the goals left to prove and the local context of assumptions. The proof engineer knows that they have proven the theorem when there are no more goals.


A \emph{proof development} in Coq consists of theorems and their associated proof scripts, potentially across multiple files. A proof engineer may even prove a series of lemmas with the sole intention of using them to prove a main theorem. Across a proof development, proof engineers often reuse bits and pieces from their proofs. However, different proofs have meaningful differences in, say, which lemmas are used. When building tools to help automate a proof engineer's proving process, it is important to fully utilize the expertise provided by the proof engineer.  
We accomplish this by directly leveraging information from existing proofs.

Let's take a closer look at the CompCert proof development, which is in \datasetname's benchmark. The file \texttt{Memdata.v} has the following theorem. 


\lstset{style=CoqStyle}

\begin{lstlisting}[numbers=none]
Lemma memval_inject_compose:
  forall f g v1 v2 v3,
  memval_inject f v1 v2 -> memval_inject g v2 v3 ->
  memval_inject (compose_meminj f g) v1 v3.
\end{lstlisting}

To prove such a theorem, one can attempt to apply tactics step-by-step, considering the current proof state and context of assumptions. Alternatively, one can instead, at each step, also take inspiration from existing proofs in the project. This is the \sysname approach. 
At each step, the \sysname tactic prediction model draws inspiration from the following proof from the file \texttt{Values.v} in the CompCert proof development, just one of the many that it retrieves.



\begin{lstlisting}[numbers=none]
Lemma val_inject_compose: forall f g v1 v2 v3, Val.inject f v1 v2 -> Val.inject g v2 v3 -> Val.inject (compose_meminj f g) v1 v3.
Proof.
  intros. inv H; auto; inv H0; auto. econstructor.
  unfold compose_meminj; rewrite H1; rewrite H3; eauto.
  rewrite Ptrofs.add_assoc. decEq. unfold Ptrofs.add. apply Ptrofs.eqm_samerepr. auto with ints.
Qed.
\end{lstlisting}

\sysname updates its sources of inspiration and knowledge at each step, as \sysname also adapts to the current proof state. The proof script for \texttt{val\_inject\_compose} would not work outright for the theorem in question, but a part of it is useful, and so \sysname begins to write the proof of \texttt{memval\_inject\_compose} as follows.

\begin{lstlisting}[numbers=none]
Proof.
  intros. inv H; inv H0; econstructor.
\end{lstlisting}

As \sysname generates the next tactic, it relies both on learned knowledge from fine-tuning and on retrieved knowledge in the form of lemmas and proofs in the project. 
Later in the proof, \sysname retrieves the following proof from earlier in \texttt{Memdata.v}.
\begin{lstlisting}[numbers=none]
Lemma memval_inject_incr: forall f g v1 v2, memval_inject f v1 v2 -> inject_incr f g -> memval_inject g v1 v2.
Proof.
  intros. inv H; econstructor. eapply val_inject_incr; eauto.
Qed.
\end{lstlisting}
\sysname also identifies \texttt{val\_inject\_compose} as a relevant helper lemma. 
It determines that it may be able to use \texttt{val\_inject\_compose} like \texttt{val\_inject\_incr} and closes the proof as follows.

\begin{lstlisting}[numbers=none]
Proof.
  intros. inv H; inv H0; econstructor. auto.
  eapply val_inject_compose; eauto.
Qed.
\end{lstlisting}

Without retrieval, a tool could get lucky and ``hallucinate'' that certain lemmas, like \texttt{val\_inject\_compose} exist. However, \sysname does not have to rely on luck because it can retrieve relevant lemmas and proofs.  

Thus, through retrieving both proofs and lemmas from the project at each step, \sysname is able to prove the theorem in question, while other state-of-the-art tools like Tactician and Proverbot are not.

\section{The \sysname Approach}
\label{sec:methodology}

\begin{figure*}
    \centering
    \includegraphics[width=\linewidth]{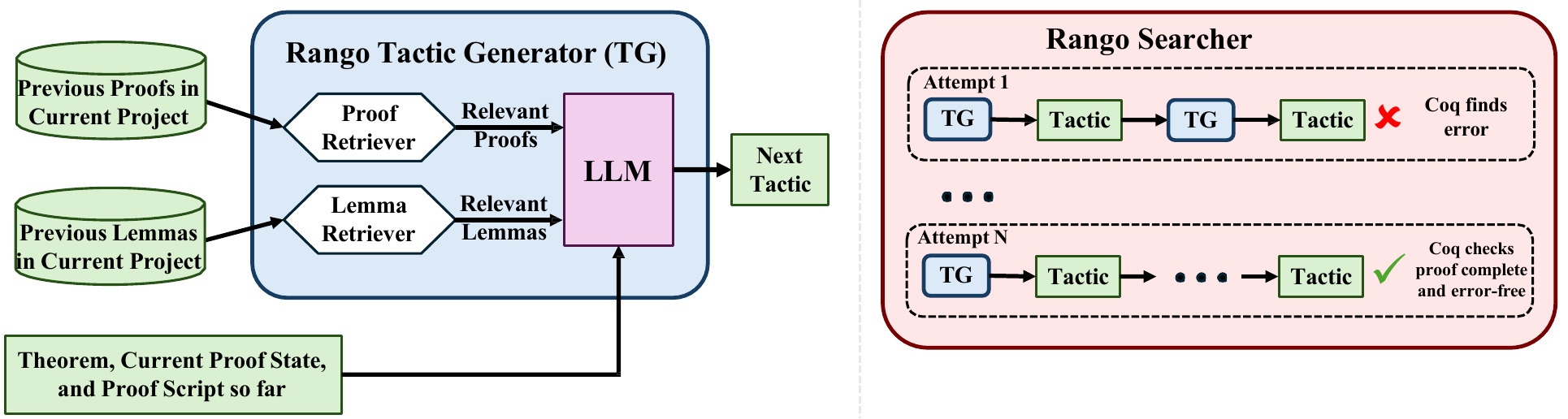}
    \caption{Overview of \sysname's architecture. \sysname's tactic generator uses retrieved relevant proofs and lemmas from the current project as input to an LLM (along with the theorem, current proof state, and proof script so far) to predict the next tactic. \sysname's searcher uses the tactic predictions to attempt to synthesize a complete proof. A proof attempt is correct if Coq determines that it has no errors and there are no more goals left to solve.}
    \label{fig:overview}
\end{figure*}

\sysname synthesizes Coq proofs using relevant proofs and lemmas from the current project at every step of the proof, adapting to the project and to the state of the proof.
To do this, \sysname uses two subsystems.
The first subsystem, the \emph{tactic generator}, generates individual proof steps, or \emph{tactics}.
The second subsystem, the \emph{searcher} (Section~\ref{sec:searcher}), uses the tactic generator to search for a complete proof by composing generated tactics. Figure~\ref{fig:overview} illustrates the interaction between the tactic generator and the searcher.

\sysname's tactic generator can be further broken down into three components.
The first two components, the \emph{proof retriever} (Section~\ref{sec:proofretriever})  and the \emph{lemma retriever} (Section~\ref{sec:lemmaretriever}), determine which proofs and lemmas, respectively, are relevant for generating the next tactic.
The third component, the \emph{language model} (Section~\ref{sec:lm}), takes as input relevant proofs, relevant lemmas, a proof script, and a proof state, and then generates the next tactic in the proof.

In the remainder of this section, we describe each of the three components in the tactic generator, and we describe the searcher.

\subsection{Proof Retriever}
\label{sec:proofretriever}
The proof retriever selects relevant previously completed proofs in the current project to provide as context to the language model as it generates the next tactic. 
At a given proof step, the proof retriever selects from a set of proofs called the \emph{proof bank}.
The \emph{proof bank} consists of proofs from earlier in the file, or from one of the file's dependencies.  
Only proofs from the current project are in the proof bank. 
At every proof step, the proof retriever selects the $k$ most relevant proofs from the proof bank for generating the next tactic.

To find the $k$ most relevant proofs at a given point in the proof, the proof retriever compares the current proof state, $s$, with proof states from proofs in the proof bank.
The proof retriever defines the relevance of a proof $P$ from the proof bank to be the maximum similarity between $s$ and a proof state $s_i$ from $P$.
Specifically, given a function \emph{similarity} that determines the similarity between two proof states,
the proof retriever defines the relevance of a proof in the proof bank as

\begin{equation*}
    \emph{relevance(P)} = \max_{s_i \in \emph{states(P)}} \emph{similarity}(s, s_i)
\end{equation*}
where \emph{states(P)} is the set of proof states in $P$.

To determine the similarity between two proof states, the proof retriever uses the BM-25 information retrieval technique~\cite{Robertson2009}.  
Given a set of documents and a query, BM-25 assigns each document a score based on its relevance to the given query.
BM-25 determines the relevance of a document by comparing the word frequencies from the document to the word frequencies from the query.
If the document and the query have similar word frequencies, then BM-25 considers the document to be relevant. 
In its calculation, BM-25 down-weighs words that appear across many documents as these words are often not helpful for determining relevance. 
Note that BM-25 is a \emph{sparse retrieval} technique because it retrieves based on word frequencies.

When the proof retriever uses BM-25, ``documents'' correspond to proof states from proofs in the proof bank, and the ``query'' corresponds to the current proof state.
The ``words'' in a proof state are the identifiers used in the proof state.
We then define \emph{similarity} in terms of BM-25 as follows: the \emph{similarity} between the current proof state $s$ and a proof state from a proof in the proof bank $s_i$ is defined to be the relevance assigned to $s_i$ by BM-25. 

Note that information retrieval techniques other than BM-25 can be substituted into \sysname's proof retriever. 
In Section~\ref{sec:proof-ret-ablation}, we explore the use of retrieval techniques that rely on neural networks, known as \emph{dense retrieval} techniques.

\subsection{Lemma Retriever}
\label{sec:lemmaretriever}
The lemma retriever retrieves the statements of the lemmas previously defined in the current project that could be directly used in the current proof. 
Note that the lemma retriever does not retrieve the proofs of these lemmas, just the statements.

The structure of the lemma retriever is similar to the structure of the proof retriever. 
The lemma retriever has access to the \emph{lemma bank}, which is defined as the set of lemmas appearing earlier in the current file, or in one of the file's dependencies.
The lemma bank only considers lemmas from the current project.

Like the proof retriever, the goal of the lemma retriever is to select the $j$ most relevant lemmas.
The lemma retriever uses the sparse retrieval algorithm TF-IDF \cite{Spark1972} to assign a relevance score to each lemma in the lemma bank with respect to the current proof state. 
Note that in this case, the ``query'' given to TF-IDF is the current proof state, the set of ``documents'' are the lemmas in the lemma bank, and the ``words'' in a lemma correspond to the identifiers in the lemma.
Again, like in the proof retriever, the lemma retriever can use information retrieval techniques besides TF-IDF.

\subsection{Language Model}
\label{sec:lm}
At a given proof step, the language model generates the next tactic using the following inputs: 
\begin{itemize}
  \item \emph{relevant proofs} retrieved from the proof retriever
  \item \emph{relevant lemmas} retrieved from the lemma retriever
  \item the \emph{theorem statement} and the \emph{proof script} thus far
  \item the current \emph{proof state}
\end{itemize}

To obtain a language model that can effectively use this information, we \emph{fine-tune} a pretrained decoder-only LLM. 
We construct fine-tuning examples from a set of training projects (see Section~\ref{sec:data}). 
Each fine-tuning example consists of a prompt containing the four inputs mentioned above, and a target containing the next tactic.
Since the language model is a decoder-only LLM, the inputs and targets are concatenated during fine-tuning. Following prior work~\cite{First23fse}, the loss is only computed over the target so that the model learns to conditionally generate the target given the input and not the input itself.

We create each training example exactly as we would during inference. 
That is, we run the proof retriever and lemma retriever at every proof step in our dataset, and then construct the prompt using the retrieved proofs and lemmas. 

Note that language models can only process a limited number of tokens.
To account for this constraint, we allocate a maximum number of tokens to each of the four inputs in each training example. 
Furthermore, we allocate a maximum number of tokens that can be generated by the language model. 
We truncate each input as follows. 
We keep the largest whole number of relevant proofs that does not exceed the token limit.
We do the same for relevant lemmas.
We keep the longest suffix of the theorem statement and proof script that does not exceed the token limit.
We keep the longest suffix of the proof state that does not exceed the token limit.
We truncate the output at training time by keeping the longest prefix that does not exceed the token limit.
At inference time, we limit the number of tokens that the model can generate.

\subsection{Searcher}
\label{sec:searcher}
\label{sec:proof-search}
Given a tactic generator, the searcher attempts to find a sequence of tactics that completes the proof.
To find a sequence of tactics, the searcher uses a procedure that we call \emph{rollout search}.
Rollout search consists of a sequence of \emph{rollouts}, where in each rollout, the searcher samples a tactic from the tactic generator using temperature sampling.
The searcher then appends the tactic to the current proof, and uses Coq to check the resulting proof attempt.
The proof attempt will be in one of three states:
complete, invalid, or incomplete.
If the proof attempt is complete, meaning that Coq shows no more goals to be proven, then the search is successful.
If the proof attempt is invalid, then the tactic sampled from the language model resulted in an error, and the searcher begins a new rollout. 
If the proof attempt is incomplete, the searcher continues the current rollout by sampling yet another tactic from the tactic generator.
The searcher executes rollouts until it finds a complete proof, or until a timeout.

\section{The \datasetname Dataset}
\label{sec:data}
As part of our work we created \datasetname, a new dataset of Coq proofs mined from open-source GitHub projects.
We collect theorems and their respective proofs from all open-source repositories that listed Coq as its primary language 
as of November 5th, 2023.
We applied no other filters to our data collection (e.g., stars or number of contributors), since Coq itself ensures that successfully compiled files provide sound proof data. 

We first attempted to automatically compile each repository by executing any existing \emph{Makefile} or by compiling each individual file in the order specified by a \emph{\_CoqProject} file present in the repository.
Then, we used CoqPyt~\cite{carrott2024coqpyt} to validate each Coq file. We considered a file to be valid if it reports no errors during compilation. We only included repositories in our training dataset with at least 1 valid file. We note that both the compilation and validation steps are executed using Coq 8.18, so files not compatible with this version are excluded.
For each valid Coq file obtained, we used CoqPyt to extract information from each proof step in the file.
For each proof step, we extracted its textual representation, corresponding proof state, and its context.
The context of a proof step includes the premises used in the proof step. 
We also use CoqPyt to collect the set of premises that are available at every point in the file.

After collecting \datasetname, we split it into a training set, a validation set, and a benchmark (test set).
We use the training set to train the language model. We use the benchmark to evaluate \sysname against other tools, and to evaluate the individual components of \sysname. We use the validation set to tune model hyper-parameters, and to experiment with different configurations of our tool. 

\datasetname's benchmark consists of all CoqGym~\cite{Yang19} projects that compiled in Coq 8.18.
CoqGym is a benchmark that was used to evaluate previous tools~\cite{Sanchez20, First20oopsla, First22icse}, for which the projects were compiled in Coq 8.9.
We also added CompCert~\cite{Leroy09} to our set of test projects to evaluate how \sysname could help formally verify real-world software.
Finally, we add projects from the Coq-Community that are committed to long-term maintenance to our benchmark and validation set.
We put any project not in our benchmark or validation set in the training set. 
Finally, to prevent against ``copy-and-pasted'' theorems, we omit files from our training set where a theorem statement exactly matched a theorem statement from the validation set or the benchmark. 

\begin{table}[t] 
    \centering
    \caption{Attributes of the training set, benchmark, and validation set.}
    \begin{tabular}{lrrrr}
    \toprule
    \multirow{2}{*}{\textbf{Attribute}} & \multicolumn{4}{c}{\textbf{Count}} \\
    \cmidrule{2-5}
    & \emph{\scriptsize{Training}} & \emph{\scriptsize{Benchmark}} & \emph{\scriptsize{Validation}} & \emph{\scriptsize{Total}} \\
    \midrule
        Repositories & 
            $2,208$ & 
            $12$ & 
            $6$  & 
            $2,226$ \\
       Theorems / Proofs & 
            181,562 & 
            10,396 & 
            4,971 & 
            196,929 \\
        Proof States / Steps & 
            2,008,543 & 
            162,989 & 
            53,983 & 
            2,225,515 \\
    \bottomrule
    \end{tabular}
    \label{tab:dataset}
\end{table}

Table~\ref{tab:dataset} shows how the relevant proof data is numerically distributed across all splits. The training set contains a substantial portion of the repositories, proofs, and proof steps in comparison to the benchmark and validation set. 
These results showcase the abundance of proof data used for training, which is guaranteed to pertain to valid Coq proofs due to the use of CoqPyt during the mining process.


\begin{figure}[t]
  \centering
  \includegraphics[width=0.95\linewidth]{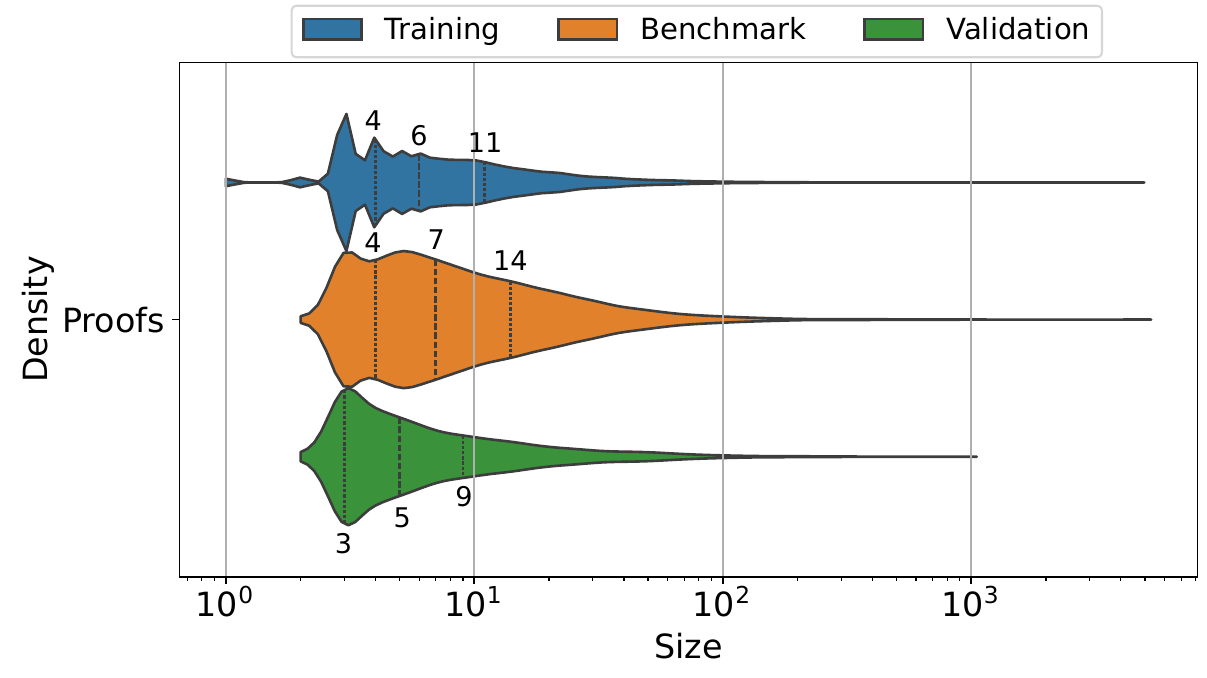}
  \caption{Violin chart over the size of proofs in the training set, benchmark, and validation set. Proof size is the number of steps in the proof. The quartiles are indicated near the corresponding dashed lines. }
  \label{fig:steps-violin}
\end{figure}

\begin{figure}[t]
  \centering
  \includegraphics[width=0.95\linewidth]{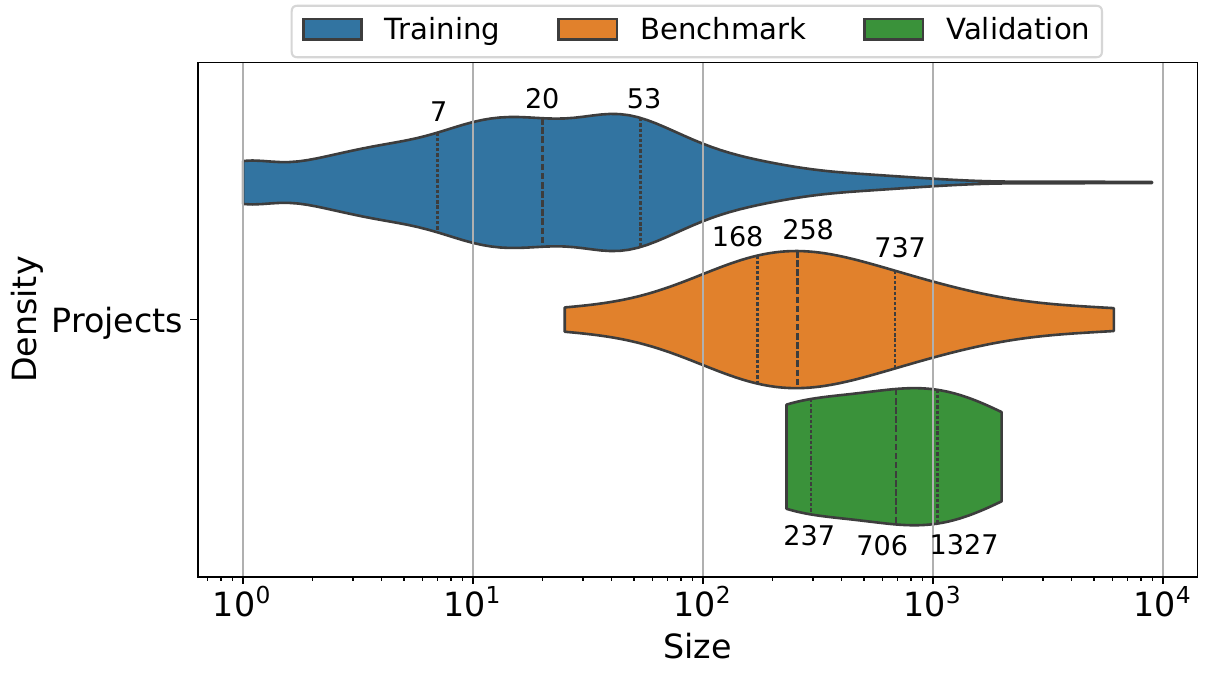}
  \caption{Violin chart over the size of projects contained in the training set, benchmark, and validation set. Project size is the number of proofs in the project. The quartiles are indicated near the corresponding dashed lines. }
  \label{fig:proofs-violin}
\end{figure}

Figures~\ref{fig:steps-violin} and~\ref{fig:proofs-violin} plot the distribution of proofs and projects, respectively, according to their size.
In terms of proof size, the number of steps per proof follows a similar distribution for all splits. 
In terms of project size, the training set contains mostly smaller projects with fewer proofs, while the reverse is true for the benchmark and validation set.
This asymmetry is justified as we intend to evaluate how the training generalizes to other, possibly larger, projects.

\section{Evaluation}

\begin{table*}[]
    \centering
    \caption{Comparison of Theorems Proven between \sysname and state-of-the-art proof synthesis tools}
    \begin{tabular}{lccccccc}
        \toprule
        \multirow{2}{*}{\textbf{Project}} & 
        \multicolumn{4}{c}{\textbf{CoqStoq Evaluation}} & 
        \multicolumn{3}{c}{\textbf{Graph2Tac Evaluation}} \\
        \cmidrule{2-8}
         & 
         \emph{\sysname} & 
         \emph{Tactician} & 
         \emph{Proverbot} &  
         \# Theorems &
         \emph{\sysname$^{\star}$} & 
         \emph{Graph2Tac${}^{\star}$} & 
         \# Theorems \\
        \midrule
        CompCert &    
            $\mathbf{1,977\ (32.5\%)}$ &
            $1,427\ (23.4\%)$ &
            $1,308\ (21.5\%)$ & 
            $\phantom{0}6,091$ & 
            --- & 
            --- & 
            --- \\
        FourColor &
            $\mathbf{\phantom{0,}212\ (15.8\%)}$ &
            $\phantom{0,}155\ (11.6\%)$ &
            $\phantom{0,}133\ (\phantom{0}9.9\%)$ &
            $\phantom{0}1,341$ & 
            --- & 
            --- & 
            --- \\
        MathClasses &
            $\mathbf{\phantom{0,}303\ (39.7\%)}$ &
            $\phantom{0,}242\ (31.7\%)$ &
            $\phantom{0,0}98\ (12.8\%)$ & 
            $\phantom{00,}763$ & 
            --- & 
            --- & 
            --- \\
        BuchBerger &
            $\mathbf{\phantom{0,}180\ (27.4\%)}$ &
            $\phantom{0,}150\ (22.8\%)$ &
            $\phantom{0,}103\ (15.7\%)$ & 
            $\phantom{00,}658$ & 
            --- & 
            --- & 
            --- \\
        RegLang &
            $\mathbf{\phantom{0,0}42\ (13.2\%)}$ &
            $\phantom{0,0}38\ (11.9\%)$ &
            $\phantom{0,0}29\ (\phantom{0}9.1\%)$ & 
            $\phantom{00,}318$ & 
            --- & 
            --- & 
            --- \\
        PolTac &
            $\mathbf{\phantom{0,}216\ (83.4\%)}$ &
            $\mathbf{\phantom{0,}216\ (83.4\%)}$ &
            $\phantom{0,}140\ (54.1\%)$ &
            $\phantom{00,}259$ & 
            $127\ (83.5\%)$ & 
            $\mathbf{138\ (90.8\%)}$ & 
            $152$ \\
        Huffman &
            $\mathbf{\phantom{0,0}82\ (32.0\%)}$ &
            $\phantom{0,0}57\ (22.3\%)$ &
            $\phantom{0,0}47\ (18.4\%)$ & 
            $\phantom{00,}256$ & 
            --- & 
            --- & 
            --- \\
        Zfc &
            $\phantom{0,0}75\ (36.2\%)$ &
            $\mathbf{\phantom{0,0}78\ (37.7\%)}$ &
            $\phantom{0,0}37\ (17.9\%)$ &
            $\phantom{00,}207$ & 
            $\mathbf{\phantom{0}75\ (36.2\%)}$ & 
            $\phantom{0}59\ (28.5\%)$ & 
            $207$ \\
        ZornsLemma &
            $\mathbf{\phantom{0,0}51\ (29.1\%)}$ &
            $\phantom{0,0}36\ (20.6\%)$ &
            $\phantom{0,0}22\ (12.6\%)$ &
            $\phantom{00,}175$ & 
            --- & 
            --- & 
            --- \\
        ExtLib &
            $\mathbf{\phantom{0,}105\ (63.6\%)}$ &
            $\phantom{0,}102\ (61.8\%)$ &
            $\phantom{0,0}32\ (19.4\%)$ & 
            $\phantom{00,}165$ & 
            --- & 
            --- & 
            --- \\
        DBLib &
            $\mathbf{\phantom{0,0}74\ (53.6\%)}$ &
            $\phantom{0,0}67\ (48.6\%)$ &
            $\phantom{0,0}51\ (37.0\%)$ &
            $\phantom{00,}138$ & 
            $\mathbf{\phantom{0}74\ (54.0\%)}$ & 
            $\phantom{0}68\ (49.6\%)$ & 
            137 \\
        HoareTut &
            $\mathbf{\phantom{0,00}8\ (32.0\%)}$ &
            $\phantom{0,00}7\ (28.0\%)$ &
            $\phantom{0,00}7\ (28.0\%)$ &
            $\phantom{00,0}25$ & 
            --- & 
            --- & 
            --- \\
        \midrule
        Total &
            $\mathbf{3,325\ (32.0\%)}$ &
            $2,575\ (24.8\%)$ &
            $2,007\ (19.3\%)$ & 
            $10,396$ & 
            $\mathbf{276\ (55.6\%)}$ & 
            $265\ (53.4\%)$ & 
            $496$ \\
        \midrule
        \sysname + Tool &
            --- & 
            $3,866\ (37.2\%)$ & 
            $3,724\ (35.8\%)$ & 
            $10,396$ & 
            --- & 
            $319\ (64.3\%)$& 
            $496$ \\
        \bottomrule
    \end{tabular}
    \label{tab:tool-eval}
\end{table*}

\begin{table}[]
    \centering
    \caption{Post Training Cutoff Comparison of Theorems Proven between \sysname and state-of-the-art proof synthesis tools}
    \begin{tabular}{lcccc}
        \toprule
         & 
         \emph{\sysname} & 
         \emph{Tactician} & 
         \emph{Proverbot} &  
         \# Theorems \\
        \midrule
        BB5 &    
            $\mathbf{186\ (38.5\%)}$ &
            $163\ (33.7\%)$ &
            $104\ (21.5\%)$&
            $\phantom{0,}483$ \\
        PnV &
            $166\ (24.1\%)$ &
            $\mathbf{168\ (24.4\%)}$& 
            $113\ (16.4\%)$& 
            $\phantom{0,}688$ \\
        \midrule
        Total &
            $\mathbf{352\ (30.1\%)}$ &
            $331\ (28.3\%)$ &
            $217\ (18.5\%)$ & 
            $1,171$\\
        \bottomrule
    \end{tabular}
    \label{tab:tool-eval-cutoff}
\end{table}

To understand \sysname's performance, we propose the following research questions:
\begin{itemize}[itemindent=0.8em]

  \item[\textbf{RQ1}:] How does \sysname compare against other proof synthesis tools in Coq? 

  \item[\textbf{RQ2}:] How do the proof retriever and the lemma retriever contribute to \sysname's ability to synthesize proofs?

  \item[\textbf{RQ3}:] How do alternative retrieval techniques perform in \sysname's proof retriever? 
  

  \item[\textbf{RQ4}:] How does \sysname perform when it is instantiated with a na{\"\i}ve retrieval algorithm?

  \item[\textbf{RQ5}:] How does \sysname's rollout search compare to best-first search?

  \item[\textbf{RQ6}:] What kinds of theorems can \sysname prove and how do the proofs generated by \sysname compare to the proofs generated by other tools? 
\end{itemize}

\subsection{Experimental Setup}
\sysname's language model is a fine-tuned version of the 1.3 billion parameter model DeepSeek-Coder~\cite{Guo2024deepseek}.
We fine-tune the LLM on a set of examples gathered from the training projects in our dataset \datasetname.
We fine-tune for 60,000 training steps with a batch size of 16 on 4 NVIDIA A100 GPUs.  
We use 2 gradient accumulation steps so that our effective batch size is 32.
We choose the checkpoint with the best loss on our validation set.
We fine-tune using LoRA~\cite{hu2021lora} and FSDP~\cite{ot2021fsdp}.
We use the Adam Optimizer~\cite{kingma2014adam} with a learning rate of $10^{-3}$.
We allocate 1,024 tokens to retrieved proofs, 512 tokens to retrieved lemmas, 512 tokens to the theorem and proof script, 1,024 tokens to the proof state, and 128 tokens to the output.

During proof synthesis, \sysname is allocated a single NVIDIA RTX 2080 for inference, and a single CPU with 16GB of RAM for proof checking.
We use a 10 minute timeout for all of our proof attempts. 
Our timeout does not include the initialization costs of loading and compiling the file. We use a temperature of 1.0 for sampling.

\subsection{\textbf{RQ1}: \sysname versus Other Tools}
\label{sec:cross-tool-eval}
We evaluate \sysname against three state-of-the-art proof synthesis tools for Coq: Proverbot9001~\cite{Sanchez20}, Tactician~\cite{Blaauwbroek20}, and Graph2Tac~\cite{Blaauwbroek24ICML}.
Proverbot9001 (which we refer to as Proverbot) uses a custom architecture involving several Gated Recurrent Units and feed-forward neural networks~\cite{Sanchez20}. 
Graph2Tac also uses a custom architecture based on Graph Neural Networks.
It uses this architecture both to predict which tactic to use next, and which definitions in the environment (including helper lemmas) should be given as an argument to the tactic. 
Therefore, Graph2Tac can retrieve helper lemmas and definitions from the environment just like \sysname.
Lastly, Tactician maintains a database of proof states which is defined similarly to \sysname's proof bank.  
At every proof step, Tactician finds the most similar proof states in its database by comparing sets of identifiers using $k$-NN~\cite{Blaauwbroek20}. 
Then, Tactician performs a search by directly applying tactics that were used at similar proof states.

We evaluate \sysname, Tactician, and Proverbot on all 12 projects from \datasetname's benchmark using Coq 8.18.
Note that neither Tactician nor Proverbot are built to utilize a GPU during their respective proof search procedures.
Therefore, we run each tool using a single CPU.
We run Tactician using a 10 minute timeout.  
Proverbot is configured to run with depth limits instead of timeouts, so for most of the reported proofs, Proverbot fails before 10 minutes. 
In Table~\ref{tab:tool-eval}, we report the results for \sysname, Tactician, and Proverbot on the $10,396$ theorems in the CoqStoq benchmark.
\sysname finds $29\%$ more proofs than Tactician, and $66\%$ more proofs than Proverbot.


In Table~\ref{tab:tool-eval}, we also report results for Graph2Tac.
Note that since Graph2Tac is only compatible with Coq 8.11, we evaluate Graph2Tac on different project versions than \sysname.
Furthermore, there are only 3 of the projects in \datasetname's benchmark that Graph2Tac was not directly trained on. 
Therefore, we can only fairly evaluate on these three projects.
We only compare theorems whose statements match exactly between project versions.
Note that this does not guarantee that a proof in one project version will translate to a proof in the other project version since other definitions in the project may have changed between versions. 
For each theorem, we ran Graph2Tac on a single NVIDIA RTX 2080 and a single CPU with a timeout of 10 mins.
Keeping differences between project versions in mind, 
\sysname proves $4\%$ more theorems than Graph2Tac.

In the last row of Table~\ref{tab:tool-eval}, we show the combined number of theorems proven between \sysname and each other tool.
We can see that each tool finds proofs for a significant subset of theorems where \sysname could not find a proof. 
Running \sysname alongside Tactician, Proverbot, and Graph2Tac leads to $16\%$, $12\%$, and $16\%$ respective increases in the number of theorems proven over running \sysname alone. 

One limitation of the \datasetname benchmark is that its projects were created before the pretraining cutoff of \sysname's underlying LLM. 
So, although we \emph{do not} fine-tune on the projects in the \datasetname benchmark, there is a risk that \sysname's underlying LLM saw them during pre-training. 
Therefore, we evaluate \sysname on two projects whose first commit on GitHub occurred after the pretraining cutoff of \sysname's underlying LLM \cite{bb5, pnv}.
The results of this evaluation are shown in Table~\ref{tab:tool-eval-cutoff}.
On these two projects, \sysname proves $6\%$ more theorems than Tactician and $62\%$ more theorems than Proverbot.
Unfortunately, we cannot compare \sysname to Graph2Tac on these projects because they do not compile with Coq 8.11.

\begin{table}[t]
    \centering
    \caption{Proof \& Lemma Retriever Ablation}
    \label{tab:retriever-ablation}
    \begin{tabular}{lc}
    \toprule
    \textbf{System} & \textbf{Theorems Proven} \\
    \midrule
       \sysname &
            $\sfrac{150}{500} = 30.0\%$ \\ 
       \sysname w/o Lemmas &    
            $\sfrac{145}{500} = 29.0\%$ \\ 
       \sysname w/o Proofs &    
            $\sfrac{102}{500} = 20.4\%$ \\ 
       \sysname w/o Retrieval &
            $\sfrac{\phantom{0}93}{500} = 18.6\%$ \\ 
    \bottomrule
    \end{tabular}
\end{table}

\begin{tcolorbox}
\textbf{Takeaway 1}: \sysname synthesizes 29\% more proofs than Tactician and 66\% more proofs than Proverbot9001 on \datasetname's benchmark.
\sysname also synthesizes 4\% more proofs than Graph2Tac on its subset of \datasetname's benchmark. 
\end{tcolorbox}

\subsection{\textbf{RQ2}: Contribution of Proof and Lemma Retrievers}
\begin{figure}
    \centering
    \includegraphics[width=0.9\linewidth]{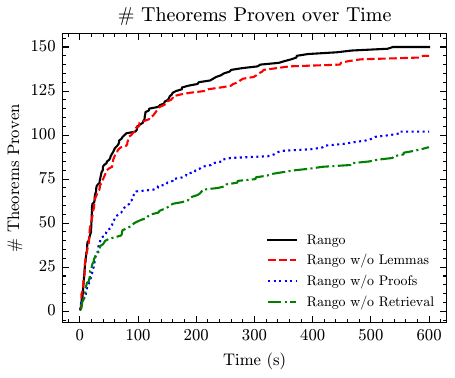}
    \caption{Number of Theorems proven by \sysname variants over time in seconds.}
    \label{fig:retriever-ablation}
\end{figure}
To determine the contribution of the proof retriever and the lemma retriever to \sysname's performance, we train one version of \sysname without the proof retriever, one version without the lemma retriever, and one version with neither the proof retriever nor the lemma retriever. 
We evaluate each of these variants on a randomly selected subset of 500 theorems from \datasetname's benchmark.
We call this subset of theorems our \emph{ablation set}.
We report the percentage of theorems that each of these variants prove in Table~\ref{tab:retriever-ablation}.

Table~\ref{tab:retriever-ablation} shows that retrieval-augmentation is essential to \sysname's success.
\sysname proves 61\% more theorems than the variant without retrieval.
Table~\ref{tab:retriever-ablation} also shows that the proof retriever contributes to \sysname's success more than the lemma retriever.
Indeed, \sysname proves 47\% more theorems than the variant without a proof retriever whereas \sysname only proves 3\% more theorems than the variant without a lemma retriever.
This is indeed one of our main contributions: we show that adding relevant \emph{proofs}, not just lemmas (as prior work had done) is important to making \emph{retrieval-augmented proving} perform better.
We show the number of proofs found by \sysname and its variants over time in Fig~\ref{fig:retriever-ablation}, which visualizes the large effect of \sysname's proof retriever.

\begin{table}[t]
    \centering
    \caption{Proof \& Lemma Retriever Ablation on a Random File-Wise Split}
    \label{tab:retriever-ablation-rnd}
    \begin{tabular}{lcc}
    \toprule
    \multirow{2}{*}{\textbf{System}} & \multicolumn{2}{c}{\textbf{Theorems Proven}} \\
    \cmidrule{2-3}
    & \textbf{\datasetname Split} & \textbf{File-Wise Split} \\
    \midrule
       \sysname &    
            $\sfrac{160}{500} = 32.0\%$ & 
            $\sfrac{184}{500} = 36.8\%$ \\ 
       \sysname w/o Lemmas &    
            $\sfrac{159}{500} = 31.8\%$ &
            $\sfrac{177}{500} = 35.4\%$ \\ 
       \sysname w/o Proofs &    
            $\sfrac{111}{500} = 22.2\%$ &
            $\sfrac{140}{500} = 28.0\%$ \\ 
       \sysname w/o Retrieval &
            $\sfrac{\phantom{0}89}{500} = 17.8\%$ &
            $\sfrac{127}{500} = 25.4\%$ \\ 
    \bottomrule
    \end{tabular}
\end{table}

We also investigate the contribution of \sysname's proof and lemma retrievers to its ability to adapt to held-out projects.
We show that \sysname's proof and lemma retrievers capture information about a project that would otherwise need to be stored in the weights of the underlying LLM. 
We show this by first training a separate version of each variant on an \textit{inter-file split} as opposed to \datasetname's inter-project split. 
That is, we randomly split the \emph{files} in the \datasetname dataset into a training, validation, and testing set. 
Then, we train each variant on the training set of the inter-file split. 
This way, the inter-file versions of the variants have information about all \datasetname projects in their weights. 
Then, we compare inter-file variants to inter-project variants on a subset of 500 theorems that are in the testing sets of \emph{both} the inter-file split and the inter-project split.
We show the results in Table~\ref{tab:retriever-ablation-rnd}.
From these results, we see that all variants benefited from having project-specific information in the weights of their underlying LLMs.
However, the variants that did not have proof retrieval benefited more than variants that did have proof retrieval.
Specifically, \sysname without proofs and \sysname without proofs and lemmas obtained increases of $26\%$ and $43\%$ in the number of theorems proven when they were trained on an inter-file split.
In contrast, \sysname and \sysname without lemmas obtained more modest increases of $15\%$ and $11\%$.
This indicates that \sysname's proof retriever captures a significant amount of the information that would be gained by training directly on files from the current project.

\begin{tcolorbox}
\textbf{Takeaway 2}:
\sysname's proof retriever and lemma retriever are imperative to its success.
Together, they contribute to a 61\% increase in the number of theorems proven, and \sysname's proof retriever alone contributes to a 47\% increase in the number of theorems proven. This demonstrates the importance of adding relevant proofs, not just lemmas as prior work had done.
\end{tcolorbox}


\subsection{\textbf{RQ3}: Comparing Retrieval Algorithms for Proof Retrieval}
\label{sec:proof-ret-ablation}
We compare \emph{sparse retrieval} techniques to \emph{dense retrieval} techniques for retrieving proofs.
Sparse retrieval techniques, such as BM-25, are those that retrieve information based on word counts.
In contrast, dense retrieval techniques retrieve information based on vector embeddings derived from a neural network. 

Note that a comparison has been made between sparse retrieval techniques and dense retrieval techniques for the selection of premises \cite{Yang23}.  
However, the techniques for training models to select premises do not transfer to training models to retrieve proofs due to a difference in objectives. 
In premise selection, the objective is to predict whether or not a premise will be used in the next proof step.
For proof retrieval, the objective is to retrieve the proofs that are most helpful for generating the next proof step. 
However, at training time, there is no way to know which proofs satisfy this objective.
Therefore, standard supervised learning techniques are not applicable.  


\begin{table}[t]
    \centering
    \caption{Proof Retrieval Variants}
    \label{tab:proof-ablation}
    \begin{tabular}{lll}
    \toprule
    \textbf{Proof Retrieval} & \textbf{Theorems Proven} \\
    \midrule
       BM-25 &
        $\sfrac{3,325}{10,396} = 32.0\%$ \\
       TF-IDF &
        $\sfrac{3,291}{10,396} = 31.7\%$ \\
       Codebert &
        $\sfrac{2,283}{10,396} = 22.0\%$ \\
    \bottomrule
    \end{tabular}
\end{table}

Recall that a main requirement of the proof retriever is the ability to compare proof states.
\sysname uses the BM-25 algorithm to determine the similarity between proof states.
Alternatively, to compare two proof states, \sysname could use a neural network to compute a vector embedding that contains semantic information about each proof state. 
Then, the similarity between two proof states could be defined as the cosine similarity between the vector embeddings.
We implemented this kind of proof retriever using the CodeBert 125M parameter model~\cite{feng2020codebert}.
We compare the proof retriever used in \sysname to a version that uses dense embeddings in Table~\ref{tab:proof-ablation}.
We also include another popular sparse retrieval algorithm, TF-IDF in Table~\ref{tab:proof-ablation}. 
Because the difference in performance between BM-25 and TF-IDF is small, we ran this ablation on the entire \datasetname benchmark instead of our ablation set to ensure the precision of our comparison.
\sysname proves $46\%$ more theorems than a variant that uses CodeBert embeddings to retrieve proofs.  
Thus, embeddings from CodeBert do not capture similarities between proof states that are relevant for proof retrieval.  
We also see that \sysname proves $1\%$ more theorems than a variant that uses TF-IDF for proof retrieval.  


\begin{tcolorbox}
\textbf{Takeaway 3:}
Using the BM-25 sparse retrieval technique for proof retrieval led to 46\% more proven theorems than using CodeBert dense embeddings and 1\% more proven theorems than using another sparse retrieval technique, TF-IDF.
\end{tcolorbox}

\subsection{\textbf{RQ4}: Rango with Na{\"\i}ve Retrieval Variant}
\label{sec:prefix-eval}


\begin{table}[]
    \centering
    \caption{Comparing \sysname, \sysname-PRE, and a Hybrid Approach}
    \begin{tabular}{lcc}
        \toprule
        \sysname Variant & 
        \textbf{\datasetname} &
        \textbf{Cutoff} \\
        \midrule
         
        \sysname & 
            $\sfrac{3,325}{10,396} = 32.0\%$ &
            $\sfrac{352}{1,171} = 30.1\%$ \\

         \sysname-PRE &
            $\sfrac{3,259}{10,396} = 31.3\%$ & 
            $\sfrac{350}{1,171} = 29.9\%$ \\
         
         \sysname-Hybrid &  
            $\sfrac{3,478}{10,396} = 33.5\%$ &
            $\sfrac{380}{1,171} = 32.5\%$ \\
        \bottomrule
    \end{tabular}
    \label{tab:prefix-eval}
\end{table}

\sysname uses a proof retriever and a lemma retriever to gather relevant context for a fine-tuned LLM to use when generating the next tactic in a proof.
A na{\"\i}ve form of retrieving proofs and lemmas could use the lines directly preceding the theorem being proven as context to the LLM.
We call this retrieval technique \emph{prefix retrieval}.
In this section, we explore \sysname-PRE: a variant of \sysname whose only retrieval mechanism is prefix retrieval.  
Table~\ref{tab:prefix-eval} shows \sysname-PRE's performance on the \datasetname benchmark, and on our two post-cutoff projects.
Note that while \sysname proves more theorems than \sysname-PRE, the margin is small. 
We know that \sysname-PRE cannot retrieve context outside of the current file. 
However, we observed that it still performed well compared to \sysname, which can retrieve proofs and lemmas throughout a project.
This led us to the following hypothesis: when the required context is in close proximity to the current theorem, \sysname-PRE is preferable since it presents this context to the LLM as it was originally written. 
However, when the required context is not in close proximity to the current theorem, \sysname is preferable since it is able to retrieve proofs and lemmas across the project.
Following this hypothesis, we created \sysname-Hybrid to capitalize on advantages from both \sysname and \sysname-PRE. 
\sysname-Hybrid simply alternates between rollouts using \sysname and \sysname-PRE. 
We can see that \sysname-Hybrid leads to a $4\%$ increase in the number of theorems proven by \sysname, and a $7\%$ increase in the number of theorems proven by \sysname-PRE.

\begin{tcolorbox}
\textbf{Takeaway 4:}
\sysname proves more theorems than \sysname-PRE. 
\sysname and \sysname-PRE can be combined into a hybrid search procedure that proves $4\%$ more theorems than \sysname, and $7\%$ more theorems than \sysname-PRE.
\end{tcolorbox}

\subsection{\textbf{RQ5}: Searcher Variants}
\begin{table}[t]
    \centering
    \caption{Searcher Variants}
    \label{tab:search-ablation}
    \begin{tabular}{lll}
    \toprule
    \textbf{Searcher} & \textbf{Theorems Proven} \\
    \midrule
       Rollout & 
            $\sfrac{150}{500} = 30.0\%$ \\
       Best-First Search (Beam) &
            $\sfrac{142}{500} = 28.4\%$\\ 
       Best-First Search (Temp) &
            $\sfrac{125}{500} = 25.0\%$ \\
    \bottomrule
    \end{tabular}
\end{table}

We also explore proof search alternatives in \sysname's searcher.
Recall from Section~\ref{sec:proof-search} that \sysname uses rollout search to search for a complete proof using its tactic generator.
One weakness of the rollout search is that it does not use previous proof attempts to inform subsequent proof attempts.
In this section, we compare rollout search to a \emph{best-first search}, which is standard in LLM proof search implementations~\cite{Polu23, Yang23, Jiang22, Han21, Jiang21}.
In our best-first search, we maintain a set of candidate proofs.
In each search step, we select the candidate with the highest score given by \sysname{}'s language model. 
Then, we use the tactic generator to generate $b$ distinct possible next tactics. 
Each tactic corresponds to a new candidate proof.
We continue the search in this way until we either find a complete proof, or the search times out.
Note that this search algorithm guarantees that each explored proof is distinct.
Finally, we do not include invalid proofs or redundant proofs \cite{Sanchez20} as candidates. 

We compare the searcher used in \sysname{}, based on rollout search to two best-first search configurations. 
In one configuration, the searcher samples tactics at each search step using beam decoding.  
In the other configuration, the searcher samples tactics using temperature sampling.
In each configuration, the searcher samples $b=4$ tactics at each search step.
Note that temperature sampling does not guarantee that the sampled tactics will be distinct. 
If the sample tactics are not distinct, we remove the duplicates.

We compare \sysname's rollout search with these two best-first search configurations on our ablation set.  
We show the results of this comparison in Table~\ref{tab:search-ablation}.
We can see from Table~\ref{tab:search-ablation} that \sysname proves $6\%$ more theorems than the configuration using beam decoding and $20\%$ more theorems than the configuration using temperature sampling.

\begin{tcolorbox}
\textbf{Takeaway 5}: Rollout search is a simple, yet effective technique for synthesizing proofs.
When used in \sysname, it synthesizes $6\%$ more theorems than best-first search. 
\end{tcolorbox}

\subsection{\textbf{RQ6}: Understanding \sysname's Proofs}

\begin{figure}
    \centering
    \includegraphics[width=0.9\linewidth]{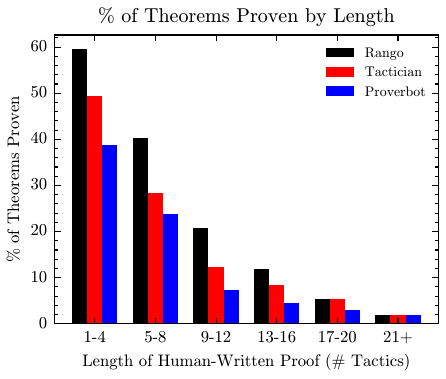}
    \caption{Percentage of theorems proven by \sysname, Tactician, and Proverbot by the length of the human-written proof.}
    \label{fig:success-by-length}
\end{figure}

In this section, we investigate which kinds of theorems \sysname can prove, and we compare the proofs generated by \sysname to the proofs generated by other proof synthesis tools. 

\subsubsection{Kinds of Theorems \sysname can Prove}
The strongest indicator that we have found for whether or not \sysname can prove a theorem is the length of its corresponding human-written proof.  
We show the success rate of \sysname, Tactician, and Proverbot by human-written proof length in Figure~\ref{fig:success-by-length}.
For all proof synthesis tools, Figure~\ref{fig:success-by-length} shows a sharp decrease in the percentage of theorems proven as the length of human-written proofs increases.

\begin{figure}
    \centering
    \includegraphics[width=0.9\linewidth]{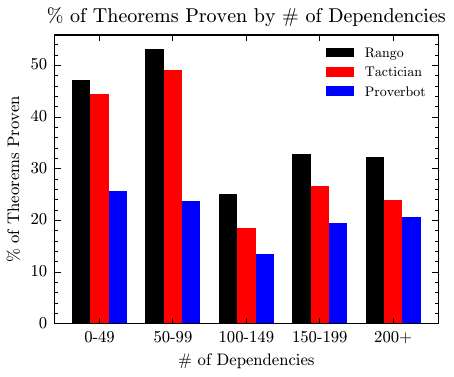}
    \caption{Percentage of theorems proven by \sysname, Tactician and Proverbot by the number of dependencies of the current file to other files in the project.}
    \label{fig:proofs-by-dep}
\end{figure}

A weaker indicator for whether or not \sysname can prove a theorem is the number of dependencies of the file containing the theorem.
We show the success rates for \sysname, Tactician and Proverbot in Fig~\ref{fig:proofs-by-dep} for files with varying numbers of dependencies where a dependency is a Coq file that is imported either directly or transitively.
We notice that the success rates for all three proof synthesis tools drop for files that have a hundred or more dependencies. 
Despite \sysname's decreased success rate on files with many dependencies, it is still able to find more proofs than other tools.  
We speculate that \sysname's retrieval mechanisms allow it to remain competitive in files with many dependencies. 
For example, consider the following theorem from the file \texttt{Values.v} in the CompCert proof development which has $300$ dependencies.

\begin{lstlisting}[numbers=none]
Theorem swap_cmpu_bool:
  forall valid_ptr c x y,
  cmpu_bool valid_ptr (swap_comparison c) x y =
    cmpu_bool valid_ptr c y x.
\end{lstlisting}

\sysname finds the following proof for this theorem using the lemma \texttt{Int.swap\_cmpu} which is defined in a different file, and is not used anywhere in \texttt{Values.v} prior to this theorem. 

\begin{lstlisting}[numbers=none]
Proof.
  destruct x; destruct y; simpl; auto. 
  rewrite Int.swap_cmpu. auto.
Qed.
\end{lstlisting}

In this case, \sysname's lemma retriever identified \texttt{Int.swap\_cmpu} as a relevant lemma for this proof, and its proof retriever found proofs that inspired this proof's structure.  

\begin{tcolorbox}
\textbf{Takeaway 6.1}:
Like other proof synthesis tools, \sysname's success-rate has a strong relationship with the length of the human-written proof.
We also found that the performance of \sysname and other proof synthesis tools decreased for files with many dependencies.
\end{tcolorbox}

\subsubsection{Qualities of Proofs Generated by \sysname}
\begin{table}[t]
    \centering
    \caption{Proof Length and Edit Distance between Machine-Generated and Human-Written Proofs}
    \label{tab:proof-edist}
    \begin{tabular}{lrrrr}
    \toprule
    \multirow{3}{*}{\textbf{System}} & 
    \multicolumn{2}{c}{\textbf{Proof Length}} &
    \multicolumn{2}{c}{\textbf{Edit Distance}} \\
    & \multicolumn{2}{c}{\textbf{in \# Tactics}} &
    \multicolumn{2}{c}{\textbf{to Human Proof}} \\
    & Mean & Median & Mean & Median \\
    \midrule
       \sysname     &  4.5  &    4  & 39.7 & 23  \\ 
       Tactician    &  4.6  &    3  & 52.8 & 37  \\ 
       Proverbot    &  6.7  &    6  & 61.2 & 48  \\
       
    \midrule 
       Human        &  4.0  &    3  & 0 & 0     \\ 
    \bottomrule
    \end{tabular}
\end{table}

To understand the qualities of proofs generated by \sysname, we measure two attributes: proof length and edit distance to the human-written proof.
We measure these attributes for \sysname, Tactician and Proverbot.
To calculate proof length, we first collected the $1,252$ theorems for which all three proof synthesis tools found proofs. 
Then, we calculated the number of tactics in each proof. 
We report the mean and median of these proof lengths in Table~\ref{tab:proof-edist}.
We also report the corresponding statistics for human-written proofs.
Human-written proofs are shorter on average than proofs generated by all proof synthesis tools.  
Proofs generated by \sysname and Tactician are of similar length, and are shorter on average than proofs generated by Proverbot.

We use string edit distance to the human-written proof as a measure of how ``human-like'' the proofs generated by \sysname are.
For each proof of the $1,252$ theorems where \sysname, Tactician, and Proverbot all found proofs, we computed the string edit distance between the proofs found by the proof synthesis tools, and their corresponding human-written proofs. 
We report these edit distances in Table~\ref{tab:proof-edist}.
On average, \sysname produces proofs that are 12 edit operations closer to human-written proofs than Tactician, and 21 edit operations closer to human-written proofs than Proverbot. 
A prior study on proof engineers found that they are constantly making repetitive repairs across their proofs due to changes to specifications or dependencies~\cite{Ringer20}. It is likely that automatically generated proofs that are in a similar style to the proof engineer's hand-written proofs would be easier for them to repair and maintain, though future work should explore this. 

\begin{tcolorbox}
\textbf{Takeaway 6.2}:
On average, the proofs synthesized by \sysname have similar or shorter lengths and smaller edit distances to human-written proofs than the proofs generated by other proof synthesis tools. 
\end{tcolorbox}

\subsection{Threats to Validity}
\label{sec:threats}

A threat to internal validity is that while the pretraining data for LLMs often consist of data from public repositories, such as GitHub, it is not publicly known what is in the pretraining data for the LLM we fine-tune for \sysname, DeepSeek-Coder 1.3B. Since \datasetname's benchmark was taken from GitHub, it is possible that it intersects with the LLM's pretraining data. Most evaluations involving LLMs suffer from this \emph{test set contamination} problem. 
We mitigate this issue by evaluating on two projects, Coq-BB5 and PnVRocqLib, that were created after the pretraining cutoff for DeepSeek-Coder 1.3B.




Another threat to internal validity is that, in our comparison of \sysname to other tools, there are some differences in the Coq versions and machines used (CPU vs GPU). 
The evaluations of \sysname, Tactician, and Proverbot all use Coq 8.18, and are all run on \datasetname's benchmark. 
However, Graph2Tac only runs in Coq 8.11. 
We evaluated \sysname and Graph2Tac on GPUs, while Tactician and Proverbot were run on CPUs since they are not intended to use GPUs. 

A threat to external validity is that while the \sysname approach could be implemented for other proof assistants, such as Isabelle and Lean, it is not known whether our results generalize across proof assistants. 
This is an interesting and important direction to be explored in future work. 

\section{Related Work}
\label{sec:related}

Formal verification aims to improve software quality, a problem that takes up
50--75\% of the total software development budgets~\cite{ODell17}. Other
methods of improving software quality include debugging~\cite{Zeller02, Johnson20icse, Brun10foser}, and
testing~\cite{Ammann08}, but only formal verification can guarantee code
correctness. Automated techniques can similarly improve program
quality~\cite{LeGoues19, Afzal21tse, Motwani23icse, Liu19, Jiang18, Zhu21,
Weiss17ase}, and can also help developers debug
manually~\cite{Eladawy24icse}, but still do not guarantee correctness, and,
in fact, often introduce new bugs~\cite{Smith15fse, Motwani22tse}.

Recent work in automating theorem proving in proof assistants, such as Coq~\cite{coq}, Lean~\cite{Moura21}, Isabelle~\cite{Nipkow02}, and Metamath~\cite{megill2019metamath}, has focused on machine-learning based approaches. Typically, with a machine learning approach, \emph{neural theorem prover} uses a model to predict the next tactic to apply, which guides a search through the space of possible proofs. Early work explored the use of LSTM~\cite{Yang19, First20oopsla, First22icse}, RNN~\cite{Huang18, Sanchez20}, and GNN-based models~\cite{Paliwal20}. Recent work has focused on the use of LLMs in neural theorem proving, either fine-tuning models~\cite{Polu20, Jiang21, Han21, Jiang22, Yang23} or prompting pretrained ones~\cite{azerbayev2023llemma, thakur2023languageagent, jiang2023draft, zheng2023lyra}. 

Large foundation models have demonstrated incredible capabilities on a wide range of tasks~\cite{palm, openai2023gpt4, touvron2023llama, team2023gemini}. To adapt to knowledge-intensive tasks and new domains, researchers have explored the use of retrieval-augmented generation (RAG)~\cite{Lewis2020} to boost performance~\cite{chen2024benchmarking}, including sparse~\cite{Spark1972, Robertson2009} and dense~\cite{karpukhin2020dense} retrieval techniques.

Recent work has explored retrieval augmentation for theorem proving, where they train models to retrieve premises, such as lemmas and definitions, that are relevant to the current proof goal, and then condition the next tactic generation on those premises~\cite{Yang23, mikuła2023magnushammer}. 
LeanDojo~\cite{Yang23} trains a model to select which premises should be included as input to its LLM tactic generation model. Magnushammer~\cite{mikuła2023magnushammer}, for Isabelle, takes this approach one step further and additionally trains a ``reranker'' model to prioritize which premises are its LLM tactic prediction model's input, though this extra step is costly. Prior work explored the use of TF-IDF for retrieving portions of premises~\cite{bansal2019learning} with the aim of training a reinforcement learning approach to theorem proving without access to human-written proofs, which our approach uses. 

Most similar to our work are Graph2Tac~\cite{Blaauwbroek24ICML} and Tactician~\cite{Blaauwbroek20}, which explore online representation learning for Coq. Graph2Tac uses GNNs to incorporate information from new definitions and theorems. Tactician uses online \emph{k}-NN to select tactics from other in-project proofs to apply in the current proof. \sysname goes beyond these approaches by being able to retrieve both new lemmas and proofs to serve as input to an LLM tactic prediction model. 

Baldur~\cite{First23fse} uses in-project information, using the lines preceding a theorem as context to its whole-proof-generation LLM. 
Section~\ref{sec:prefix-eval} showed that \sysname can be instantiated with this retrieval mechanism, and that a hybrid search procedure between \sysname and \sysname-PRE leads to the best results on the \datasetname benchmark. 
LEGO-Prover~\cite{wanglego} builds a skill library of lemmas and theorems while proving so that it may retrieve new skills learned instead of relying on a fixed library. 

Neural theorem provers have also been shown to make use of other sources of information. 
Deepseek-Prover~\cite{xin2024deepseek} learns from synthetic data.
TrialMaster~\cite{an2024learn} learns from trial-and-error paths.
Baldur~\cite{First23fse} learns from proof assistant error messages.
Passport~\cite{Sanchez-Stern23toplas} explores the use of rich identifier information. 
QEDCartographer~\cite{Sanchez-Stern25icse} uses reinforcement learning to estimate progress toward a complete proof to improve search during proof synthesis.
Autoformalization techniques~\cite{wu2022autoformalization, Azerbayev22} are guided by informal proofs in their construction of formal proofs. 

TacticToe~\cite{Gauthier17} employs an A* search. Evariste~\cite{lample2022hypertree} uses a Monte-Carlo tree search to outperform a best-first search approach in Lean. TacticZero~\cite{Wu21} learns proof search
strategies, not just tactics, through deep reinforcement learning for HOL4. 

Hammers, such as CoqHammer~\cite{Czajka18} and Sledgehammer~\cite{Paulson23}, are automation techniques used in proof assistants that also perform premise retrieval. They employ SMT solvers, like Z3~\cite{deMoura08}, and perform efficient automated reasoning to iteratively apply the set of available facts. 

In languages like F* that allow for SMT-assisted proof oriented programming, new work has shown promise in using RAG when synthesizing whole programs~\cite{chakraborty2024towards}. Other work uses RAG to generate and summarize Java and Python code~\cite{parvez2021retrieval,li2021editsum}. RAG has also been shown to be useful for program repair~\cite{wang2023rap,nashid2023retrieval}.

Our work has focused on proving properties, which is complementary to
specifying them, e.g., by generating formal specification from natural
language~\cite{Endres24, Goffi16, Mirchev24, Motwani19icse, Zhai20}. Formal languages
can be extended to be more expressive, to capture privacy
properties~\cite{Cortier17}, data-based properties~\cite{Muslu15issta,
Muslu13ni-fse}, fairness properties~\cite{Galhotra17fse, Brun18fse-nier},
among others. Some of these kinds of properties can be automatically verified
probabilistically~\cite{Thomas19science, Hoag23icse-demo, Metevier19neurips,
Giguere22iclr, Albarghouthi17}.

\section{Contributions}
\label{sec:contributions}
We developed an adaptive retrieval-augmented proving approach and tool, \sysname, to synthesize proofs for formal software verification.
\sysname improves on prior retrieval-augmented proving approaches by retrieving \emph{proofs} in addition to lemmas.
\sysname employs its retrieval mechanisms at every step in the proof to obtain the most relevant proofs and lemmas for the current proof state. 
To train \sysname, we collected \datasetname, a new dataset of Coq proofs, which includes both our training data and a curated benchmark of well-maintained projects. 
\datasetname, mined from 2,226 open-source GitHub projects, contains 196,929 theorems, their respective proofs, and 2,225,515 proof steps.
\sysname proves $32.0\%$ of theorems on \datasetname's benchmark, which is $29\%$ more than Tactician, a prior state of the art tool.
\sysname's proof retrieval is important to its success, leading to a $47\%$ increase in the number of theorems proven. 
Overall, our research shows that retrieval augmentation using in-project proofs in addition to premises is a powerful technique that can increase the proving power of automated proof synthesis tools, reducing the costs of formal verification.


\section*{Acknowledgments}
This work is supported by the National Science Foundation under grants no.\ CCF-1955457, CCF-2210243, and CCF-2220892,
by the Air Force Research Laboratory (AFRL) and Defense Advanced Research Projects Agencies (DARPA) under Contract No.\ FA8750-24-C-B044,
by DARPA under Contract No.\ HR0011-24-2-0307, 
and by FCT, Funda{\c{c}}{\~{a}}o para a Ci{\^{e}}ncia e a Tecnologia under grant BD/04736/2023 and project UIDB/50021/2020 (DOI: 10.54499/UIDB/50021/2020).

\balance
\bibliographystyle{abbrv-doi}
\bibliography{softeng,laser}

\end{document}